# Open-bore vertical MRI scanners generate significantly less RF heating around deep brain stimulation leads compared to horizontal scanners

Jasmine Vu[1,2*], Bhumi Bhusal[2*], Bach T Nguyen[2], Pia Sanpitak[1,2], Elizabeth Nowac[3], Julie Pilitsis[4], Joshua Rosenow[5], Laleh Golestanirad[1,2]

[1]Department of Biomedical Engineering, McCormick School of Engineering, Northwestern University, Evanston, IL, USA

[2]Department of Radiology, Feinberg School of Medicine, Northwestern University, Chicago, IL, USA

[3]Illinois Bone and Joint Institute (IBJI), Wilmette, Illinois, USA

[4]Department of Neurosciences & Experimental Therapeutics, Albany Medical College, Albany, NY, USA

[5]Department of Neurosurgery, Feinberg School of Medicine, Northwestern University, Chicago, IL, USA

*Jasmine Vu and Bhumi Bhusal contributed equally to this work.

**Corresponding Author:** Laleh Golestanirad

Email: laleh.rad1@northwestern.edu

Phone: (312) 694-1374

Address: 737 N Michigan Ave, Suite 1600

Chicago, IL 60611

## Abstract

### Objectives

Studies that assess magnetic resonance imaging (MRI) induced radiofrequency (RF) heating of the tissue in the presence of an active electronic implant have been mostly performed in horizontal, closed-bore scanners. Vertical, open-bore MRI systems have a 90° rotated magnet and generate a fundamentally different RF field distribution in the body, yet little is known about the RF heating of elongated implants such as deep brain stimulation (DBS) systems in this class of scanners. Here, we investigated whether RF-induced heating of DBS devices was significantly different in a 1.2 T vertical, open-bore MRI scanner compared to a 1.5 T horizontal, closed-bore MRI scanner.

**Materials and Methods**

In this phantom study, RF heating around the lead of a commercial DBS system implanted in an anthropomorphic phantom was evaluated in a 1.2 T vertical open-bore scanner (Oasis, Fujifilm Healthcare) and a 1.5 T horizontal closed-bore scanner (Aera, Siemens Healthineers). DBS devices were implanted following realistic lead trajectories which were extracted from computed tomography images of 30 patients. Additionally, electromagnetic simulations were performed to further assess the specific absorption rate (SAR) of RF energy in the tissue around leads with different internal structures. RF heating was compared between the two scanners using a one-tailed Wilcoxon signed rank test.

**Results**

When controlling for $B_1^+{}_{rms}$, temperature increase around the DBS lead-tip was significantly lower during RF exposure at the 1.2 T vertical, open-bore scanner compared to the 1.5 T horizontal scanner (0.36 ± 0.24 °C vs. 4.19 ± 2.29 °C, p-value = 9.1 x $10^{-7}$). Electromagnetic simulations demonstrated up to a 14-fold reduction in the maximum 0.1g-averaged SAR deposited in the tissue surrounding the lead-tip in a 1.2 T vertical MRI scanner compared to a 1.5 T horizontal MRI scanner for leads with straight and helical internal wires.

**Conclusions**

Radiofrequency experiments and electromagnetic simulations demonstrated consistently lower RF heating and power deposition around the DBS lead-tip at the 1.2 T vertical scanner compared to the 1.5 T horizontal scanner. Our simulation results suggest that this trend in heating may potentially extend to leads from other manufacturers.

**Key words**: magnetic resonance imaging, implants, safety, heating, deep brain stimulation, vertical

**Introduction**

Magnetic resonance imaging (MRI) provides excellent soft tissue contrast and clear visualization of fine anatomical structures without exposing the patient to ionizing radiation. For patients with active implantable medical devices (AIMDs) such as those with deep brain stimulation (DBS) implants, routine clinical MRI protocols are not as readily accessible due to the safety risk of radiofrequency (RF) heating in the tissue surrounding the DBS lead-tip. This is troubling, as up to 75% of patients with DBS systems will require an MRI exam during the lifetime of the device.[1]

Excessive local heating in the tissue surrounding the tips of AIMD leads arises due to the antenna effect, where the electric field of the MRI transmit coil couples with the long leads and induces currents on the lead wires. In response, the specific absorption rate (SAR) of RF energy in the tissue surrounding the lead's tip amplifies.[2–7] Through this phenomenon, patients have sustained thermal injuries and permanent neurological changes.[8,9] Consequently, DBS manufacturers have developed restrictive guidelines for MRI of patients with implanted DBS systems. For example, MR-conditional DBS systems from Abbott Medical limit MRI exams to those performed in horizontal, closed-bore scanners at a 1.5 T magnetic field strength and with a $B_1^+{}_{rms} < 1.1$ μT.[10] However, complying with these restrictions has proven to be difficult as clinical protocols that are optimized to visualize DBS targets or those in routine cardiac and musculoskeletal imaging have $B_1^+{}_{rms}$ values that far exceed these limits (see Table, Supplemental Digital Content 1, example routine sequences).

To date, the majority of studies that have assessed RF heating of AIMDs have been performed in horizontal, closed-bore MRI scanners. Vertical, open-bore systems have a 90° rotated magnet and a fundamentally different RF coil geometry which produces a notably different field distribution within the human body. However, little is known about the RF heating of AIMDs in this class of scanners. In recent simulation studies with simplified lead models, we showed that the local 1g-averaged SAR around tips of simple wire models that followed typical DBS lead trajectories was lower in a vertical, open-bore scanner compared to a conventional, horizontal scanner.[11,12] Here, we report results of the first experimental study that compares measured RF heating of a fully implanted commercial DBS device with patient-derived trajectories in a 1.2 T vertical, open-bore scanner (OASIS, Fujifilm Healthcare, Tokyo, Japan) with that of a 1.5 T conventional, horizontal scanner (Aera, Siemens Healthineers, Erlangen, Germany). Experiments were performed by replicating DBS device configurations observed in patients who underwent DBS surgery at Northwestern Memorial Hospital (NMH) and Albany Medical Center (AMC). Further, we performed electromagnetic (EM) simulations with various lead structures to assess whether experimental results of this study could potentially extend to DBS devices from other manufacturers.

**Materials and Methods**

*Replicating patient-derived DBS lead trajectories and in vivo scenarios*

It is established that RF heating of an elongated implant (such as leads in cardiovascular and neuromodulation devices) is largely affected by the implant's position within the human body and its orientation with respect to MRI electric fields.[6,13–17] Therefore, studies that aim to assess RF heating of leads should ideally do so by replicating patient-derived device configurations in an environment that mimics the *in vivo* scenario. To do this, we first identified clinically relevant DBS lead trajectories from postoperative computed tomography (CT) images of 30 patients who underwent DBS surgery in our institutions from May 2017 to September 2020. The retrospective use of patients' imaging data for the purpose of modeling and simulation was approved by NMH and AMC institutional review boards.

Lead trajectories were segmented from CT images using 3D Slicer 4.10.2 (http://slicer.org) and reconstructed in the CAD tool Rhino 6.0 (Robert McNeel & Associates, Seattle, WA) to create models of each lead trajectory. These models were 3D-printed to serve as guides to accurately position a commercial DBS lead along different trajectories (Fig.1). Once the leads were positioned in place, the guides were removed from the phantom so that their presence did not affect the heating experiments.

To provide a more realistic replication of the electric field distribution around the implanted leads, we used a multi-material anthropomorphic phantom consisting of a 3D-printed body-shaped container and a refillable skull structure. The phantom design was based on CT images of a patient with a DBS device. Details of the phantom design and construction are described elsewhere.[18] In brief, the skull was filled with a tissue mimicking gel ($\sigma = 0.40\ S/m$, $\varepsilon_r = 79$, similar to values reported for brain tissue),[19] prepared by mixing 32 g/L of edible agar (Landor Trading Company, gel strength 900 g/cm$^2$) with saline solution (2.25 gNaCL/L). The remaining head-torso component of the phantom was filled with 18 L of saline solution ($\sigma = 0.50\ S/m$) mimicking the conductivity of the average tissue. Using an agar-based solution to fill the skull was advantageous compared to using polyacrylamide gel in this study, because once cooled down to the room temperature, the agar-based solution formed a semi-solid gel which prevented movement of the leads following insertion in the skull. The thermal conductivity of the solidified agar gel was ~0.56 J/K-S[20] which was similar to that of grey matter.[21]

To further assess the degree to which RF exposure of DBS devices implanted in the anthropomorphic phantom represented the *in vivo* scenario, we performed EM simulations to calculate the distribution of MRI-induced electric fields on various coronal planes inside the phantom and compared them with electric fields inside a heterogenous human body model consisting of 32 tissue classes. Results demonstrated a good agreement between the two cases (see Figure, Supplemental Digital Content 2, electric field distributions) ensuring that the experimental results in the anthropomorphic phantom were a reliable indicator of what happens in patients. This is important as recent studies have highlighted that the electric fields (and by proxy, SAR and RF heating) inside the thin, box-shaped ASTM phantom can significantly differ from fields that are induced in the human body.[22]

*RF Heating Experiments*

A full commercial DBS system from Abbott (Abbott Medical, Plano, TX) consisting of a 40 cm lead (model 6173), a 50 cm extension (model 6371), and an implantable pulse generator (IPG) (Infinity 6660) was implanted in the anthropomorphic phantom. The DBS lead was inserted into the skull with fluoroptic temperature probes securely attached to the two most distal electrode contacts. Placement of the lead and temperature probes inside the skull emulated the location and angle of insertion for targeting the subthalamic nucleus (STN). The lead was connected to the extension and the IPG with the extension routed laterally along the neck and the IPG placed in the pectoral region.

RF heating experiments were performed at a 1.5 T Siemens Aera horizontal scanner (Siemens Healthineers, Erlangen, Germany) and at a 1.2 T Hitachi Oasis vertical scanner (Fujifilm Healthcare, Tokyo, Japan) (Fig. 2) using the body transmit coils at both scanners. The phantom was placed in the head-first, supine position, and an imaging landmark at the level of the DBS lead-tip was selected for all experiments. RF exposure was generated using high-SAR turbo spin echo (TSE) and fast spin echo (FSE) sequences such that the $B_1^+{}_{rms}$ was 4 $\mu$T at both scanners

(Table 1). Each experimental configuration included only one DBS lead with a single lead trajectory, representing cases of unilateral DBS. Temperature rise during RF exposure was measured at the DBS lead-tip using temperature probes (OSENSA, BC, Canada). The temperature was recorded continuously throughout the RF exposure for the total acquisition time (TA) of 224 seconds. The maximum temperature rise ($\Delta T_{max}$) was quantified as the difference between the baseline temperature at the onset of RF exposure and the highest measured temperature. The setup was allotted ample time to return to the baseline temperature prior to evaluating the next lead trajectory.

Although temperature recordings that are performed to ensure patient safety are usually performed for a longer RF exposure time (e.g., 15 minutes), our goal here was to test the hypothesis that RF heating was significantly lower in one scanner compared to the other one. The 224-second scan duration allowed us to maximize the number of experiments to increase the statistical power while still providing adequate time for the RF heating generated by the OASIS scanner to reach a plateau (thus allowing us to draw conclusive inferences).

*Investigating the effect of lead's length and internal structure*

To assess whether the results of our experiments could potentially extend to other models of DBS devices, we performed EM simulations with leads of varying internal structures. This allowed us to examine if the difference observed in the RF heating was specific to the electrical length of the lead in our experiments, or if it was a trend that could be observed for leads of shorter and/or longer lengths. This is important to consider because RF heating of an elongated implant is known to be a resonance phenomenon which depends on the length of the implant.[23,24] DBS leads from different manufactures consist of internal helical wires that are wound at different pitches, and thus have different electrical lengths even when the total length of the lead appears to be the same for different lead models.

*Simulation setup*

Electromagnetic (EM) simulations were implemented in ANSYS Electronic Desktop 2021 R1 HFSS (ANSYS, Canonsburg, PA). We recreated a lead trajectory that generated a large difference in $\Delta T_{max}$ between RF exposures in the horizontal versus vertical scanners (Fig 3). The coordinates along the lead trajectory were extracted during image segmentation and were used to reconstruct the model of the DBS lead. The lead and extension consisted of a core wire made of platinum-iridium ($\sigma = 4 \times 10^6 S/m$) embedded within a urethane insulation ($\sigma = 0\ S/m,\ \varepsilon_r = 3.5$). The total length of the modeled lead and extension was 90 cm to match the commercial device used during experiments; however, the electrical length of the core wire was changed by modeling either a straight wire or helical wires with pitches of 1 and 2 mm. The full DBS system was implanted in a standard homogeneous model of the human body truncated at the abdomen ($\sigma = 0.40\ S/m,\ \varepsilon_r = 79$) where a triangulated surface model of the patient's head and the DBS system were manually aligned to the standard body model via rigid transformation/registration (6 degrees of freedom) to place the device in the correct anatomical position.

A numerical model of a high-pass, radial planar 12-rung birdcage coil with the specifications of the body coil in the Oasis scanner tuned to 50.4 MHz was constructed as in our previous work.[11,12] Similarly, a high-pass 16-rung horizontal birdcage coil tuned to 63.6 MHz was constructed based on the details provided by Siemens. Both coils provided quadrature excitation with ports separated by 90°; the vertical coil had four ports while the horizontal coil had two ports. The input voltage applied to each port was adjusted to generate a mean $B_1^+$ of 4 $\mu$T on a central transverse plane passing through the center of the coil (Fig. 4).

Power deposition in the tissue adjacent to the DBS lead-tip was quantified using the SAR calculation module incorporated in HFSS. The maximum of the 0.1g-averaged SAR, $0.1gSAR_{max}$, was calculated in a (20 mm)³ cubic tissue region surrounding the lead-tip. A fine mesh resolution was applied during the simulations; the maximum tetrahedral mesh edge length was 2 mm for the tissue region around the DBS lead-tip and 0.5 mm for the DBS lead. Numerical convergence was ensured by imposing a constraint on the maximum variation of the scattering parameters between two consecutive iterations as described in our previous work.[25]

*Statistical Analysis*

A one-tailed Wilcoxon signed rank test was conducted to assess the difference in the measured $\Delta T_{max}$ between RF exposure experiments in 1.2 T vertical and 1.5 T horizontal scanners. Statistical significance was established for $p < 0.05$. Statistical analysis was performed in MATLAB 2020b (The MathWorks Inc., Natick, MA).

**Results**

*Demographics of Patients*

In total, this study included DBS lead trajectories from 30 patients (21 men) with a mean age ± standard deviation of 60.4 ± 13.5 years (Table 2). The most common DBS indication and target were Parkinson's disease (PD) and the STN, respectively.

*Experimental Temperature Measurements*

Thirty unique, clinically relevant trajectories were replicated during experiments at the 1.2 T Oasis vertical and 1.5 T Aera horizontal scanners. For all trajectories, the lead was implanted to target the right STN; 27 trajectories were contralateral to the IPG (IPG in the left pectoral region) while 3 trajectories were ipsilateral to the IPG.

Across all the trajectories, $\Delta T_{max}$ was significantly lower (p = 9.1 x 10$^{-7}$) for RF exposure at the vertical scanner compared to RF exposure at the conventional horizontal scanner for the input power of both scanners adjusted to generate the $B_1^+{}_{rms}$ of 4 µT. At the vertical scanner, the range of $\Delta T_{max}$ was 0.04 – 1.01 °C with a mean ± standard deviation of 0.36 ± 0.24 °C while the range of $\Delta T_{max}$ was 1.84 – 12.92 °C at the horizontal scanner with a mean ± standard deviation of 4.19 ± 2.29 °C (Fig. 5). Figure 5A illustrates the temperature profiles for the different lead trajectories throughout the duration of the MR sequences at the two scanners.

*Simulated Power Deposition in Tissue*

Electromagnetic simulations were performed with one example of a DBS lead trajectory that demonstrated a large difference between $\Delta T_{max}$ measured at the vertical and horizontal scanners. For the selected lead trajectory, we evaluated the effect of the lead's internal geometry on the power deposition in the tissue surrounding the DBS lead-tip. The electrical lengths of the inner conductors were 90 cm for the straight wire, 116 cm for the helical wire with a pitch of 2 mm, and 171 cm for the helical wire with a pitch of 1 mm. Figure 6 shows the distribution of 0.1gSAR$_{max}$ on an axial plane intersecting the tip of the DBS lead for the 1.2 T Oasis coil compared to the 1.5 T Aera coil across the different internal wire geometries. For a mean $B_1^+$ = 4 µT generated over an axial plane at the center of the imaging region, the range of 0.1gSAR$_{max}$ was 182.0 – 274.6 W/kg for the vertical coil compared to 2368.4 – 2949.9 W/kg for the horizontal coil. The mean ± standard deviation of 0.1gSAR$_{max}$ was 220.4 ± 48.3 W/kg and 2593.4 ± 312.3 W/kg for the vertical and horizontal coils, respectively.

**Discussion**

Current MRI guidelines for internalized DBS systems require the use of horizontal, closed-bore scanners, limiting potential applications of MRI in a continuously expanding population of patients with DBS. The safety risk of localized RF-induced tissue heating is a well-known barrier to MRI for patients with DBS implants. Recently, increased engineering efforts have targeted this problem through alteration of the design and methodologies of DBS and MRI. For example, MRI hardware modification has been proposed to reduce the antenna effect by shaping the electric field of the scanner through parallel transmit[17,26–29] or reconfigurable MRI technology.[30–33] Changes in DBS lead material[34,35] and modification of surgical lead implantation[6] have also been explored as alternative approaches. Although promising in theory, none of these techniques have found their way to the clinic yet, and efforts to enable implant-friendly MRI are ongoing.

RF heating of elongated implants (such as leads in electronic devices) in an MRI environment is the result of coupling of the transmit electric field of the MRI RF coil with conductive wires of the implant. The efficiency of this coupling is directly affected by the magnitude of the electric field[36] as well as the orientation of the E-field vector with respect to the wire.[16,23] Since vertical, open-bore MRI scanners have a 90° rotated RF coil, the E-field induced in the human body is substantially different from that of conventional solenoidal birdcage coils. Recent simulation studies showed that vertical scanners could generate lower SAR around DBS lead models compared to conventional scanners; however, these results have not been rigorously examined in experiments with patient-derived lead trajectories. In this study, we measured the RF heating of a commercial DBS system implanted in an anthropomorphic phantom following 30 unique patient-derived lead trajectories in the 1.2 T Oasis vertical scanner compared to the 1.5 T Aera horizontal scanner. The measured $\Delta T_{max}$ was reduced by 11-folds on average at the vertical scanner compared to the horizontal scanner (p = 9.1 x 10$^{-7}$). For a high SAR sequence ($B_1^+{}_{rms}$ = 4

$\mu$T), temperature increase was well below 2 °C for all trajectories at the vertical scanner whereas $\Delta T_{\max}$ up to 12°C was recorded at the horizontal scanner.

Additionally, we performed numerical simulations with leads of different internal geometries (and hence different electrical lengths) to investigate whether the results of our experiments could potentially extend to leads from other manufacturers. Correspondingly, internal wires of most DBS leads have a helical structure, both to increase mechanical flexibility and as a strategy to increase the electric inductance which can ultimately reduce MR-induced RF currents.[37,38] This means that the electrical length of internal wires is usually different from the apparent length of the lead (i.e., internal wires of a 40-cm DBS lead are much longer than 40 cm). For this reason, leads from different manufacturers—or even different lead models from the same manufacturer—do not necessarily behave similarly in an MRI environment. Our simulations with lead lengths of 90 - 171 cm showed that in all these cases, the vertical open-bore scanner generated a substantially lower SAR at the lead's tip compared to the horizonal closed-bore scanner, suggesting that our experimental results here could potentially generalize to DBS devices from other manufacturers.

In summary, we demonstrate that RF exposure from a vertical MR scanner induces significantly less heating than in a conventional, horizontal scanner. Our experimental results show that measured temperature increase did not exceed the regulatory limit of 2 °C even when a high-power sequence was applied. Similarly, simulation results suggest that the benefits of vertical MRI for reducing RF heating may apply to other DBS models than the one used in this study.

**Acknowledgment**

This work was supported by NIH grants R03EB029587 and T32EB025766.

**Table 1: MRI Pulse Sequences**

| Sequence Parameter | T1W TSE at 1.5 T Aera | T2W FSE at 1.2 T Oasis |
|---|---|---|
| TE (ms) | 96 | 96 |
| TR (ms) | 2780 | 2728 |
| Matrix size | 512 x 512 | 512 x 512 |
| Acquisition Time (sec) | 224 | 224 |
| $B_1^+{}_{RMS}$ ($\mu T$) | 4 | 4 |

Note. —FSE = fast spin echo, TE = echo time, TR = repetition time, TSE = turbo spin echo, T1W = T1-weighted, T2W = T2-weighted

**Table 2: Summary of Patient Population**

| Parameter | Value |
|---|---|
| Number of patients | 30 |
| Mean Age (y) | 60.4 ± 13.5 (21-76) |
|     Women | 57. 0 ± 22.0 (21 – 76) |
|     Men | 61.9 ± 7.8 (43 – 71) |
| Sex | |
|     Women | 9 |
|     Men | 21 |
| DBS Indication | |
|     Parkinson's disease | 21 |
|     Parkinson's disease and dystonia | 2 |
|     Dystonia | 1 |
|     Cervical dystonia | 1 |
|     Essential tremor | 1 |
|     Essential tremor and Parkinson's disease | 1 |
|     Orthostatic tremor | 1 |
|     Obsessive compulsive disorder | 2 |
| DBS target | |
|     Subthalamic nucleus | 21 |
|     Globus pallidus internus | 5 |
|     Ventral intermediate nucleus of the thalamus | 2 |
|     Ventral capsule/ventral striatum | 1 |
|     Anterior limb of the internal capsule | 1 |

Note. — Unless otherwise specified, data are number of participants.

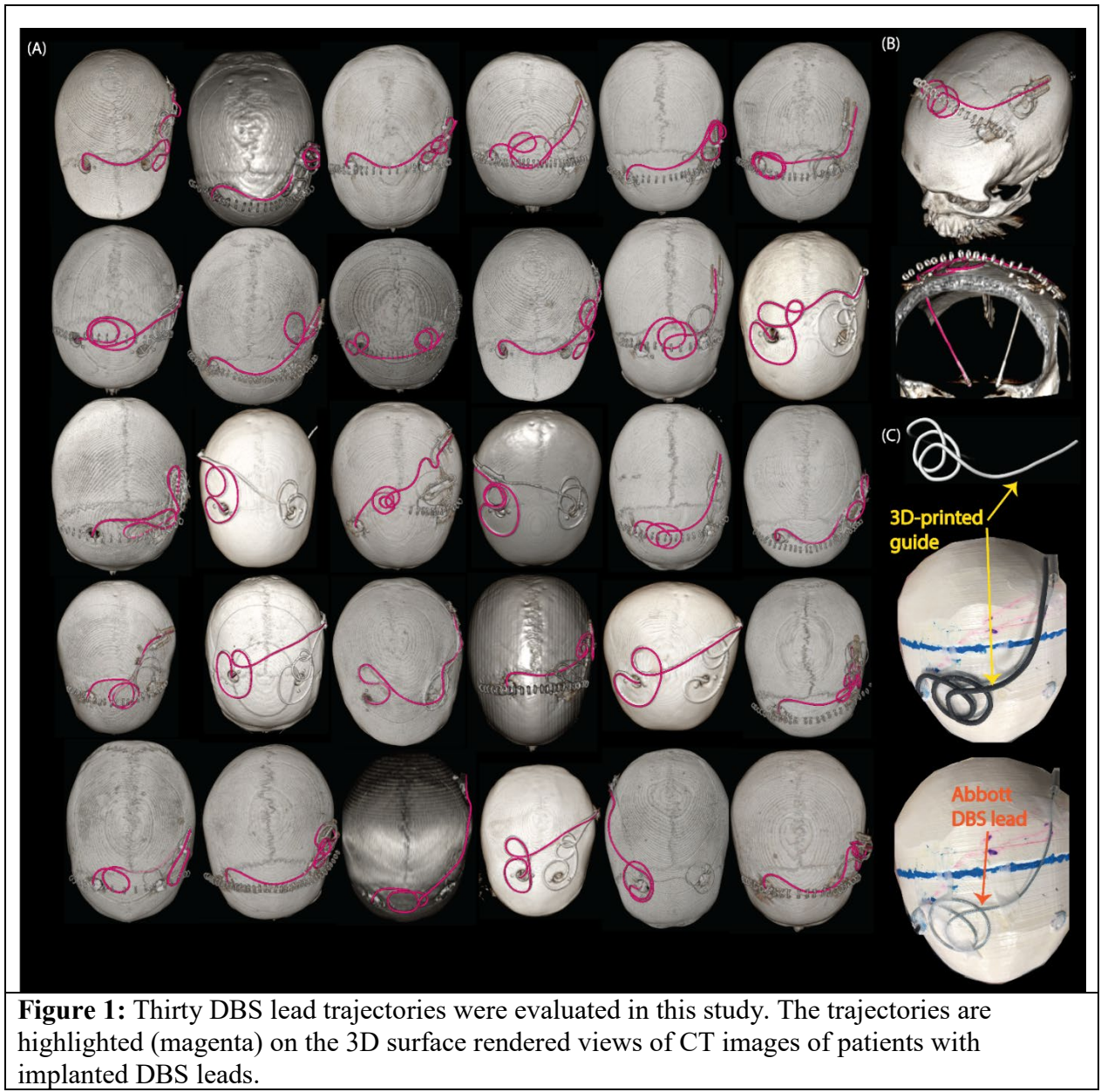

**Figure 1:** Thirty DBS lead trajectories were evaluated in this study. The trajectories are highlighted (magenta) on the 3D surface rendered views of CT images of patients with implanted DBS leads.

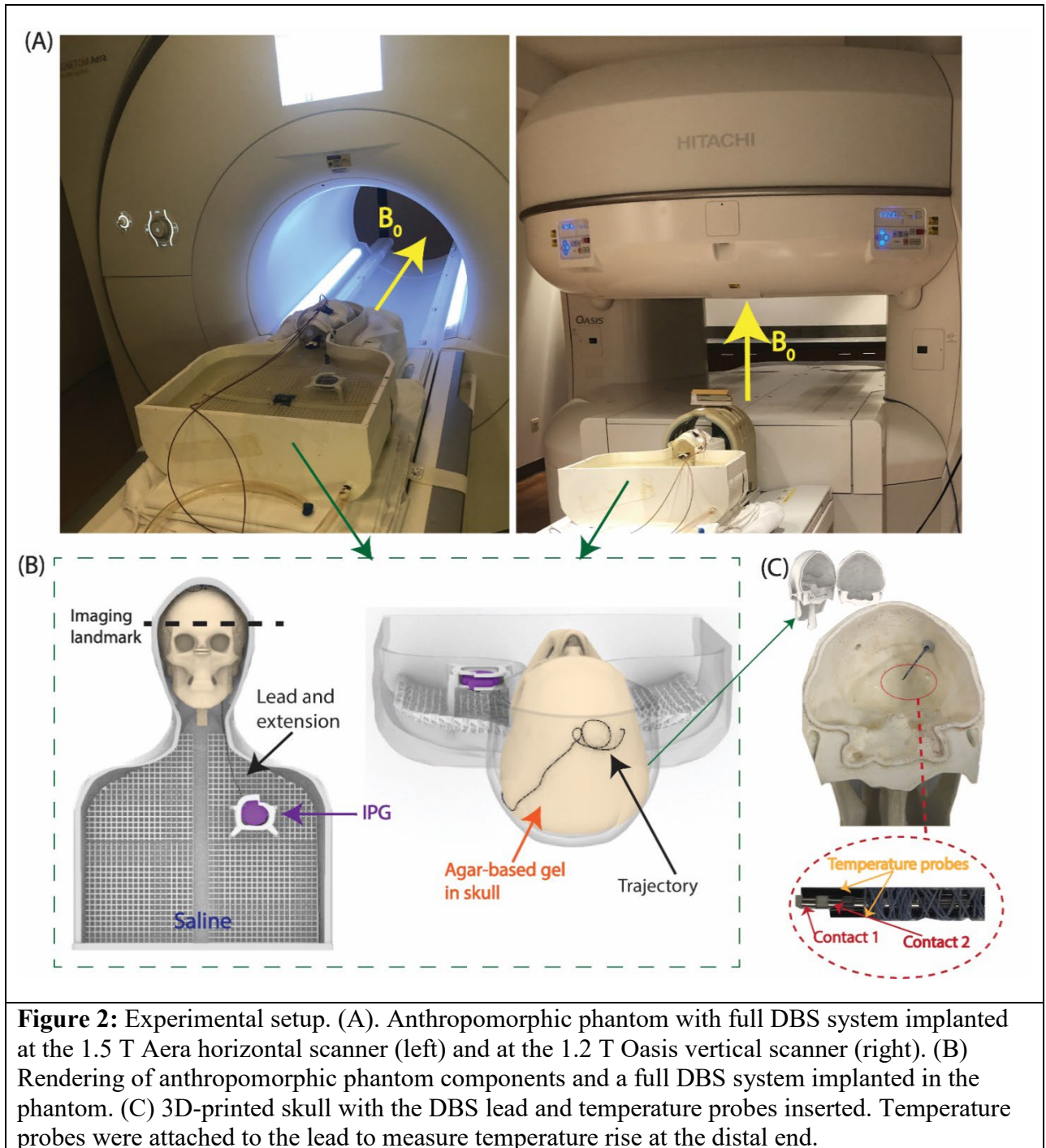

**Figure 2:** Experimental setup. (A). Anthropomorphic phantom with full DBS system implanted at the 1.5 T Aera horizontal scanner (left) and at the 1.2 T Oasis vertical scanner (right). (B) Rendering of anthropomorphic phantom components and a full DBS system implanted in the phantom. (C) 3D-printed skull with the DBS lead and temperature probes inserted. Temperature probes were attached to the lead to measure temperature rise at the distal end.

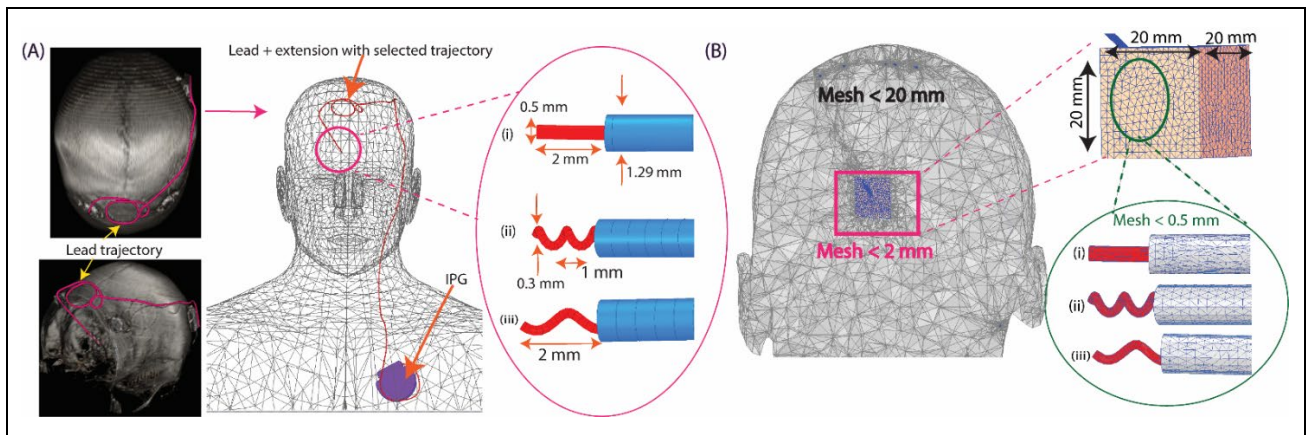

**Figure 3:** DBS system modeling and simulation setup. (A) Segmentation of the DBS lead trajectory (magenta) from the lead artifact displayed in the 3D surface rendered view of a patient's CT image and reconstruction of a full DBS model oriented in a homogeneous body model truncated at the abdomen. Reconstructed DBS lead models of the same trajectory (Pt-Ir core wire (red) within a urethane insulation (blue)) with various internal geometries were evaluated: (i) straight wire, (ii) helical wire with a 1 mm pitch, and (iii) helical wire with a 2 mm pitch. (B) Example mesh distributions of the body, tissue region for SAR calculations, and the lead.

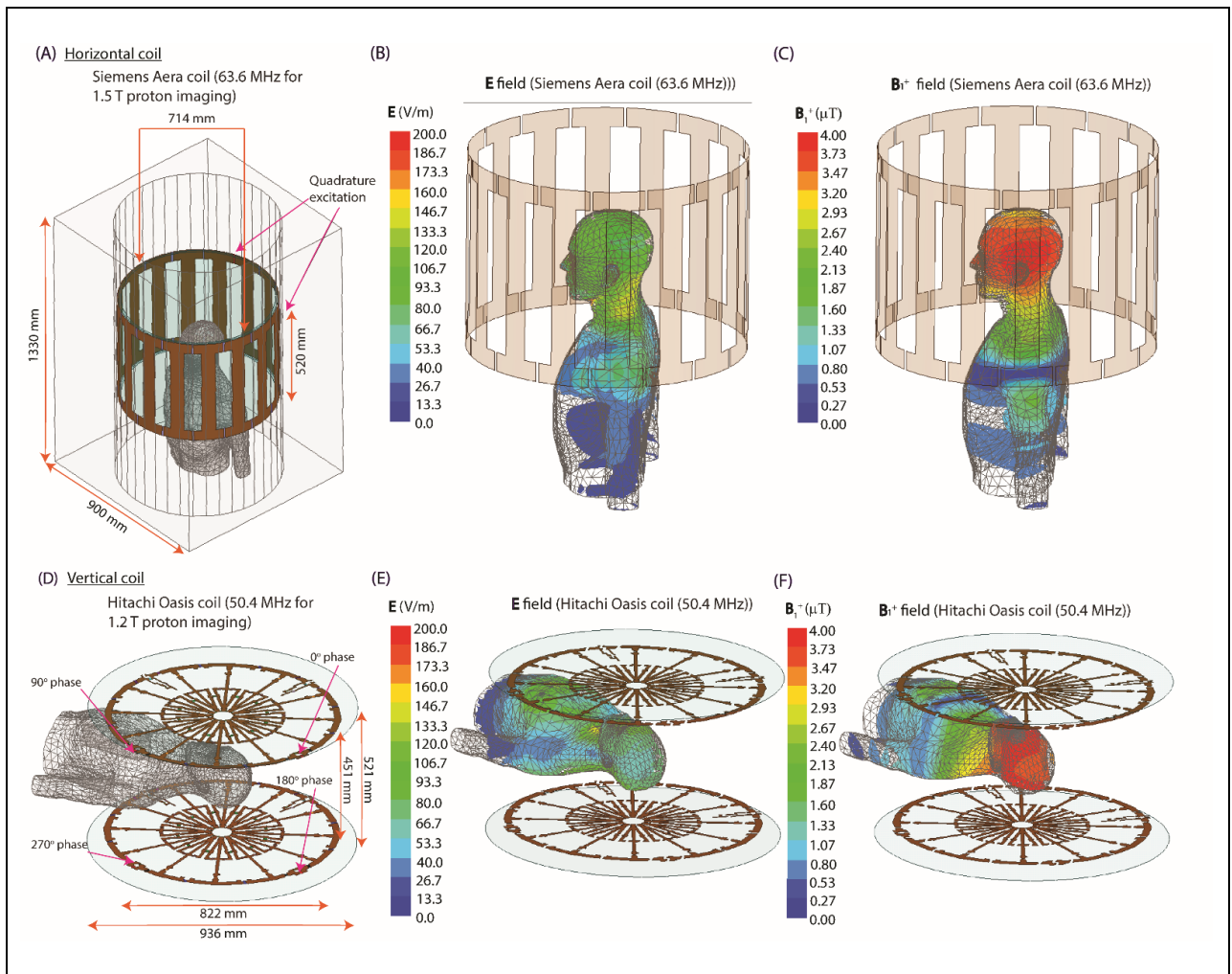

**Figure 4:** (A) Numerical models of the 1.5 T Aera horizontal birdcage coil and (D) the 1.2 T Oasis radial planar birdcage coil. The electric and $B_1^+$ field distributions in the human body in the absence of any implant when loaded in the horizontal (B, C) and vertical (E, F) coils. The mean $B_1^+$ at the isocenter of both coils was 4 $\mu$T.

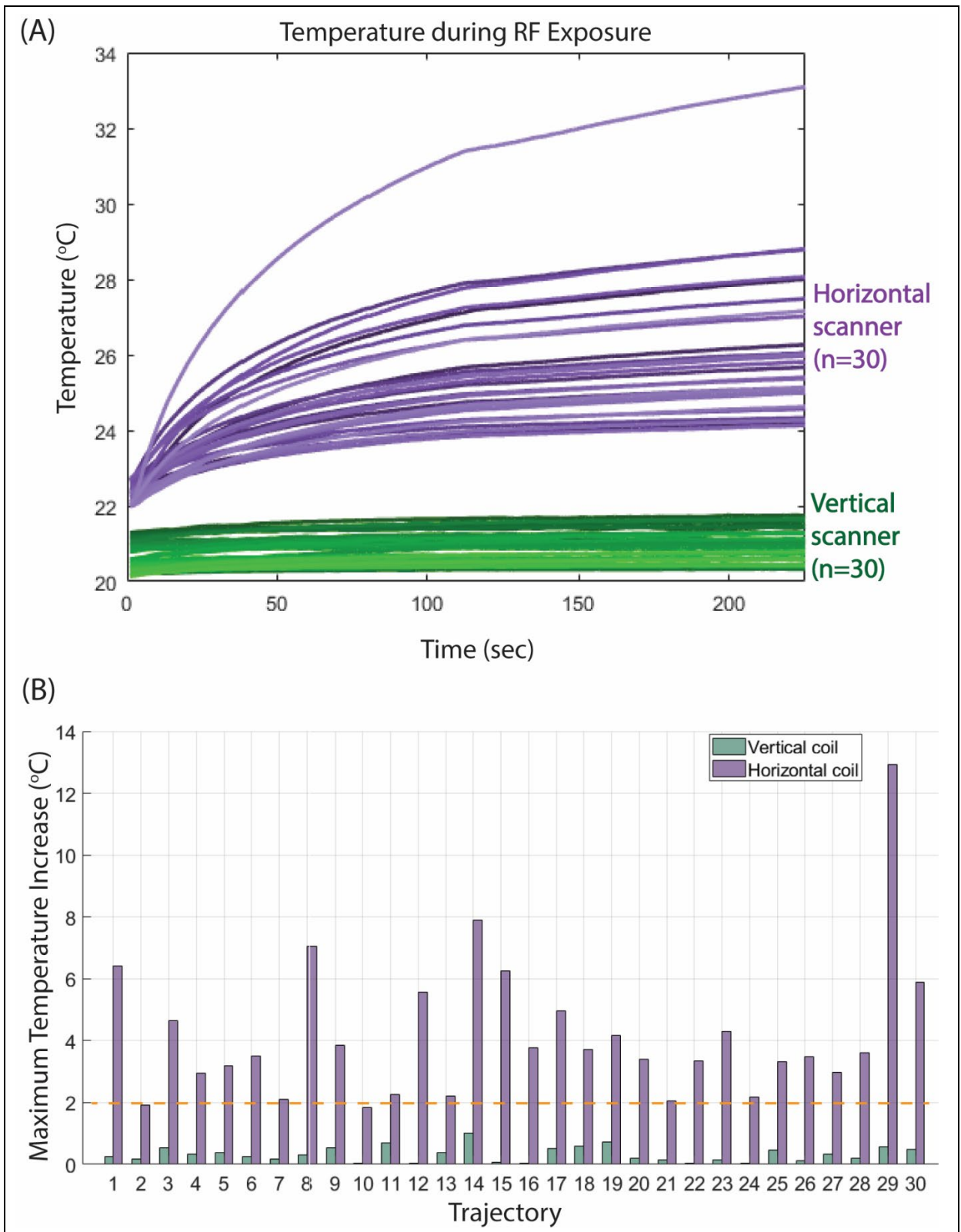

**Figure 5:** (A) Measured temperature at the DBS lead-tip during RF exposure at the vertical 1.2 T and horizontal 1.5 T scanners. (B) The maximum temperature rise during RF heating experiments for each trajectory.

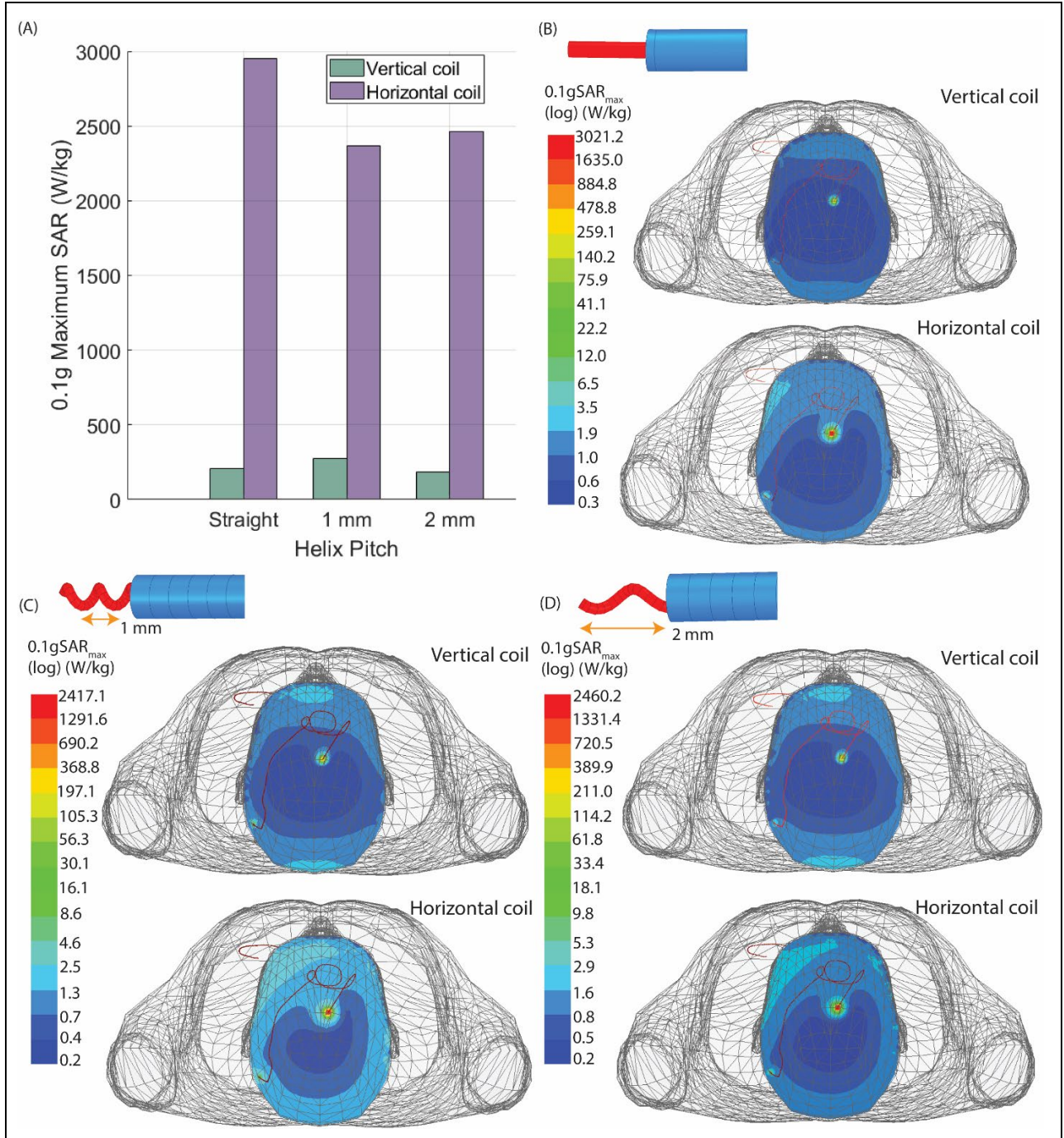

**Figure 6:** (A) Plot of simulated $0.1gSAR_{max}$ at the DBS lead-tip for wires with different electrical lengths in the vertical and horizontal coils. Spatial distribution of the simulated $0.1gSAR_{max}$ at the level of the DBS lead-tip with the internal core of the lead represented as (B) a straight wire, (C) a helical wire with a 1 mm pitch, and (D) a helical wire with a 2 mm pitch. The mean $B_1^+$ at the isocenter of both coils was 4 $\mu$T.

# Supplemental Digital Content

Supplemental Digital Content 1

**Table:** Examples of routine MRI sequence parameters and scanner-reported $B_1^+{}_{rms}$ values for different imaging protocols. Data is taken from a clinical 1.5 T scanner at Northwestern Memorial Hospital.

| 1.5 T Siemens Avanto | | | | | |
|---|---|---|---|---|---|
| **Head Landmark** | | | | | |
| **Protocol** | **TE (ms)** | **TR (ms)** | **TA (min)** | **FA (°)** | **$B_1^+$ (µT)** |
| SE T1 SAG | 10 | 450 | 2:09 | 90 | 3.54 |
| AX FLAIR | 86 | 9000 | 3:02 | 150 | 2 |
| AX T1 SE | 12 | 450 | 2:26 | 150 | 3.04 |
| AX T2 TSE | 94 | 3800 | 2:22 | 150 | 3.71 |
| **Chest Landmark** | | | | | |
| **Protocol** | **TE (ms)** | **TR (ms)** | **TA (min)** | **FA (°)** | **B1+ (µT)** |
| COR TRUFI | 1.21 | 491.33 | 0:20 | 64 | 4.01 |
| AX TRUFI | 1.18 | 328.38 | 0:13 | 64 | 3.99 |
| AX VIBE | 2.39, 4.77 | 6.9 | 0:20 | 10 | 1.49 |
| TRUFI CINE 2C | 1.16 | 38.22 | 0:06 | 60 | 3.98 |
| TRUFI CINE SA | 1.16 | 40.96 | 1:00 | 60 | 3.99 |
| **Abdomen Landmark** | | | | | |
| **Protocol** | **TE (ms)** | **TR (ms)** | **TA (min)** | **FA (°)** | **B1+ (µT)** |
| T2 HASTE COR MBH | 91 | 1400 | 0:48 | 180 | 4.2 |
| T2 HASTE FS TRA MBH | 94 | 1400 | 0:05 | 160 | 3.1 |
| T2 BLADE FS TRA | 91 | 2200 | 2:35 | 160 | 3.35 |
| T2 TSE FS TRA MBH | 86 | 4780 | 0:54 | 160 | 3.77 |

Supplemental Digital Content 2

Simulations were performed with the anthropomorphic phantom—consisting of brain tissue-mimicking material, the skull, and saline—used in RF heating experiments and a heterogeneous human body model from ANSYS consisting of 32 tissue classes (Fig. 1 and 2). Simulations were performed with both vendor-specific transmit body coils for the 1.5 T Siemens Aera scanner and the 1.2 T Hitachi Oasis scanner. These simulations were conducted in the absence of implanted DBS systems. The E-field distributions at three planes were calculated from simulations (Fig. 3 and 4). In all simulations, the input voltage was adjusted to produce a mean $B_1^+$ of 4 µT on an axial plane passing through the center of head. The E-field distributions were similar between the heterogeneous body model and our anthropomorphic phantom, demonstrating that the anthropomorphic phantom is a good representation of *in vivo* scenarios.

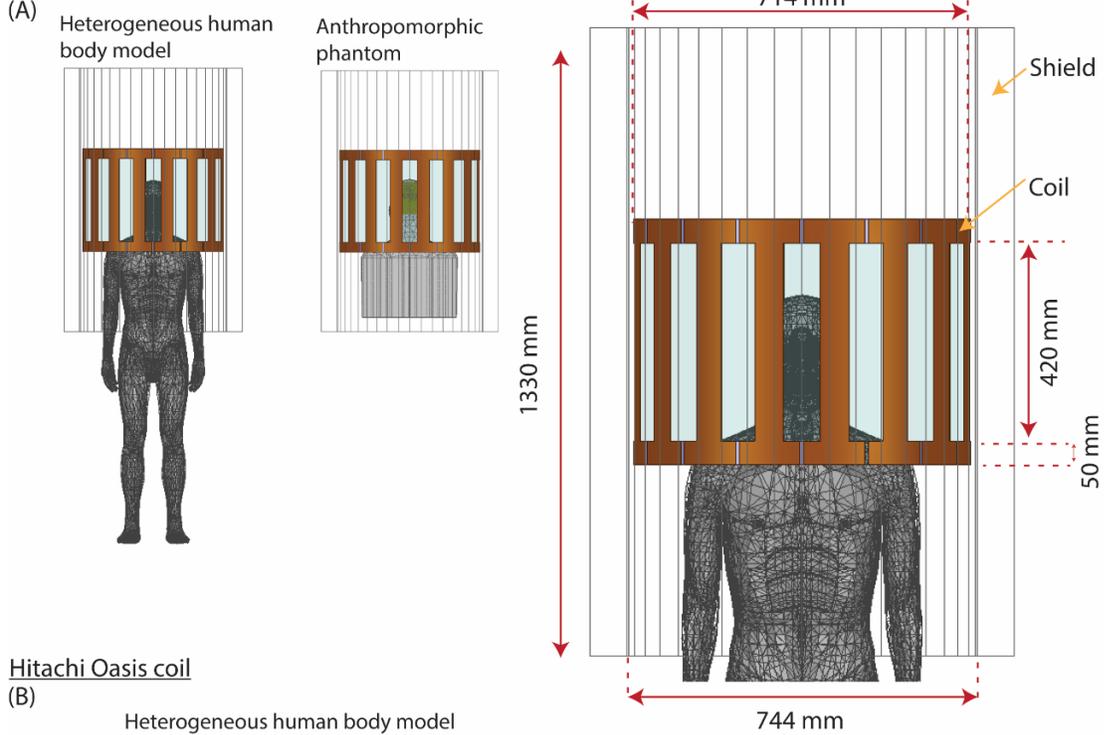
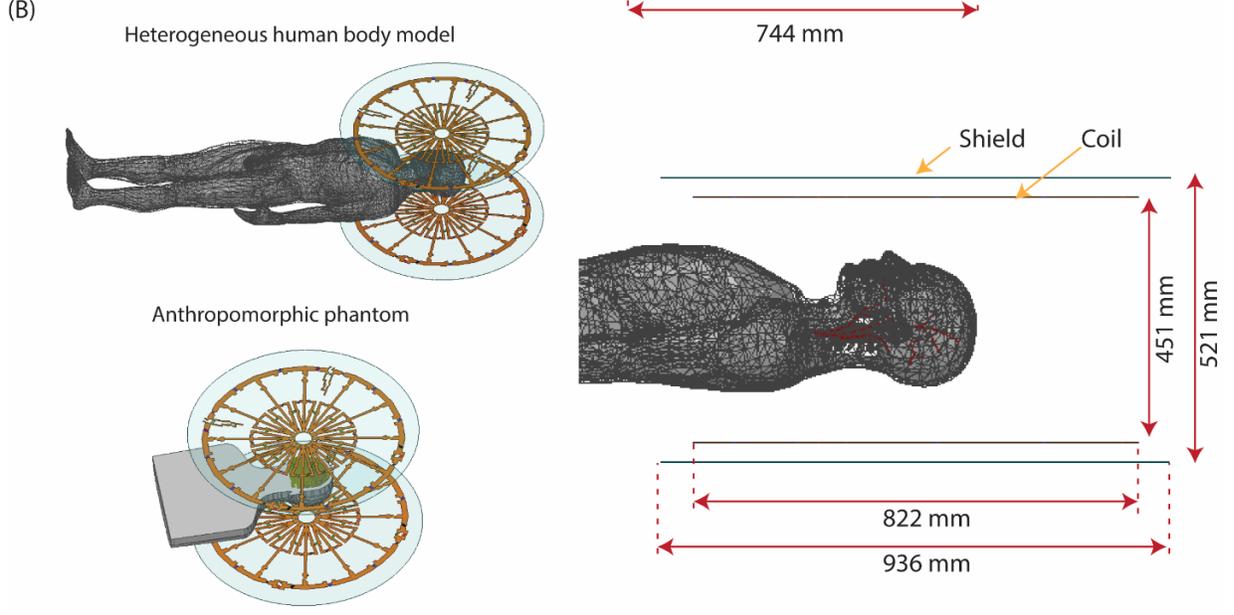

Figure 1: (A) Simulation setup of the heterogeneous human body model and the anthropomorphic phantom in the 1.5 T Siemens Aera horizontal coil. The coil and shield dimensions are also provided. (B) Simulation setup of the heterogeneous human body model and the anthropomorphic phantom in the 1.2 T Hitachi Oasis vertical coil. The coil and shield dimensions are also provided.

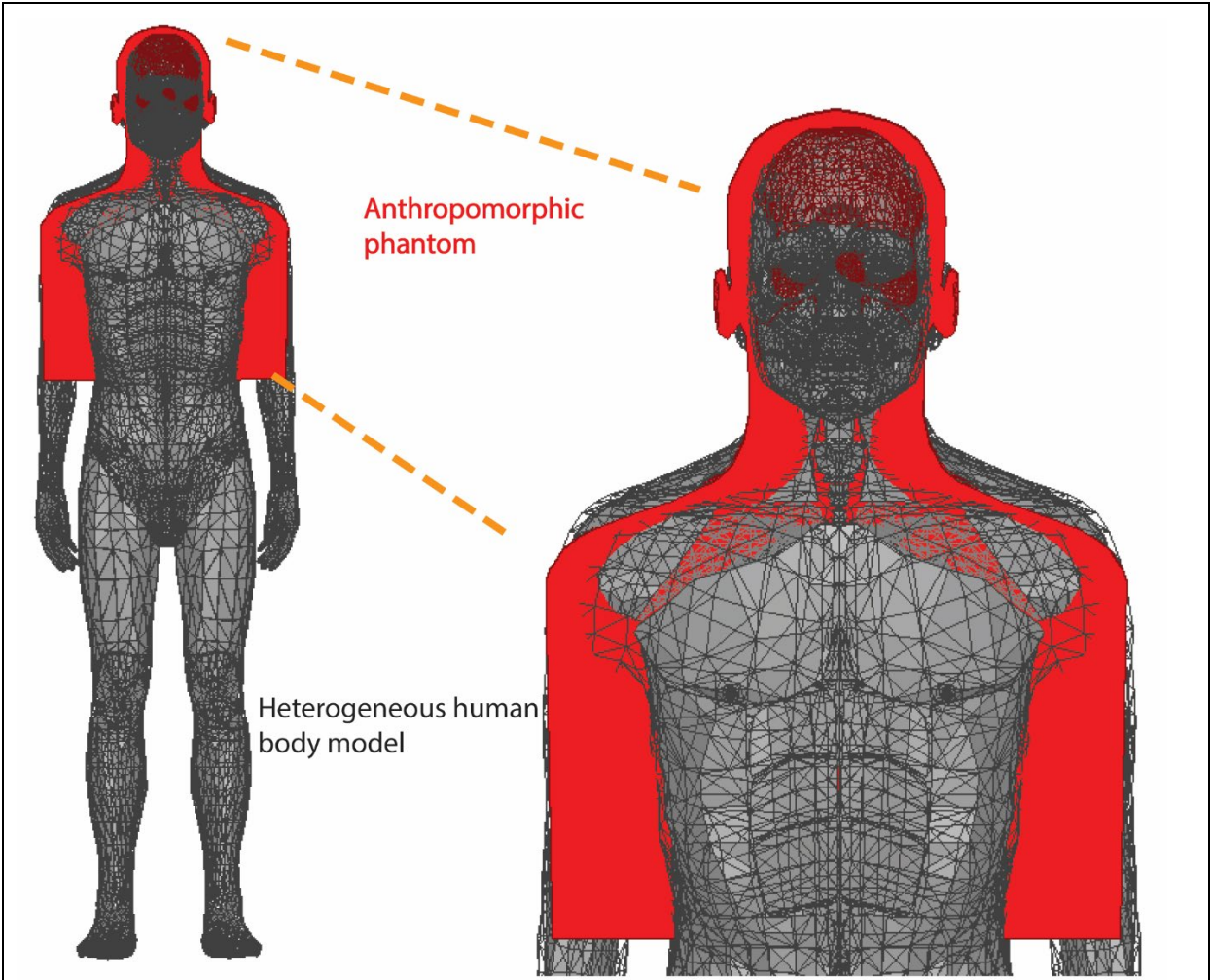

Figure 2: Comparison of the size of the anthropomorphic phantom used in experiments versus the heterogeneous human body model.

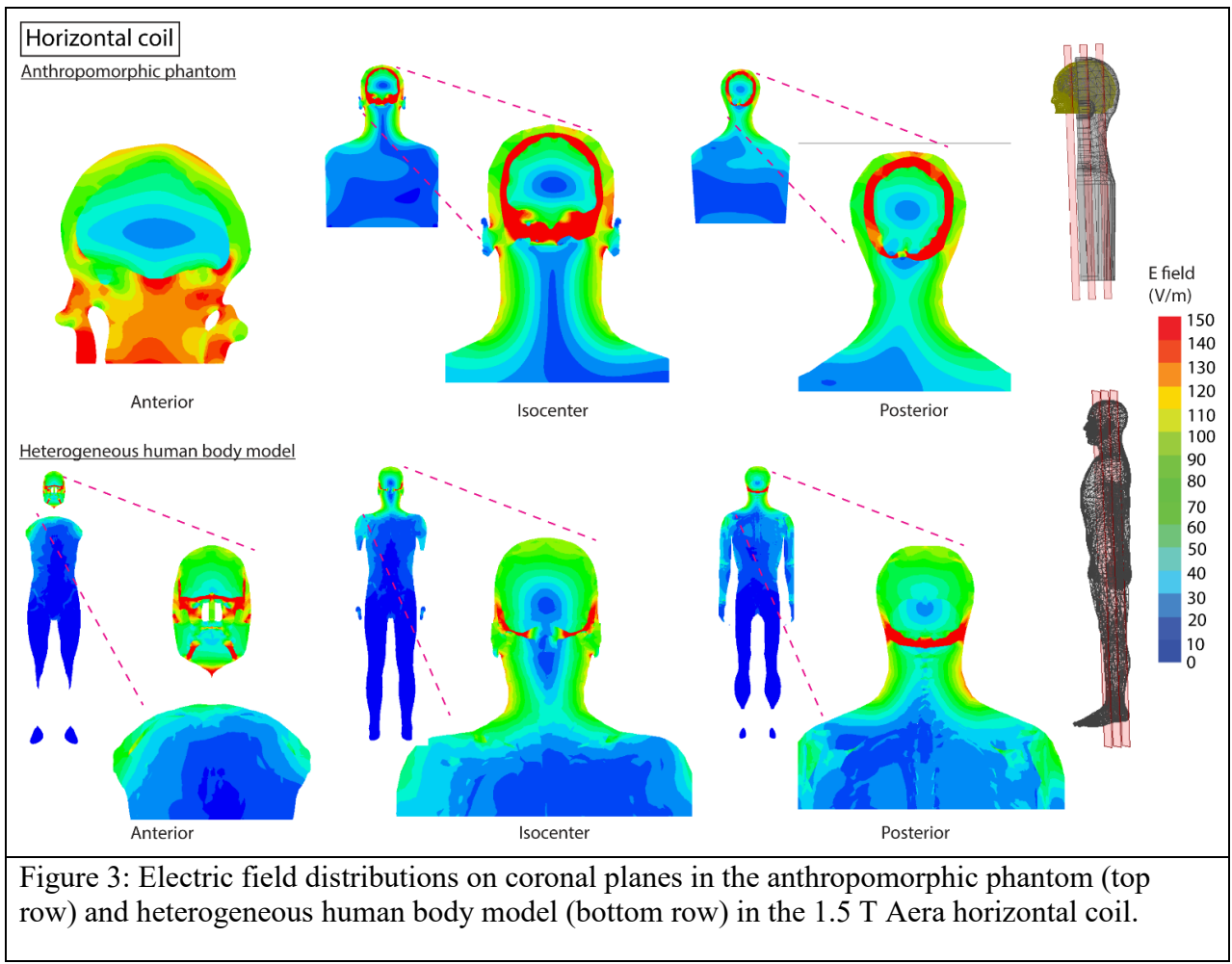

Figure 3: Electric field distributions on coronal planes in the anthropomorphic phantom (top row) and heterogeneous human body model (bottom row) in the 1.5 T Aera horizontal coil.

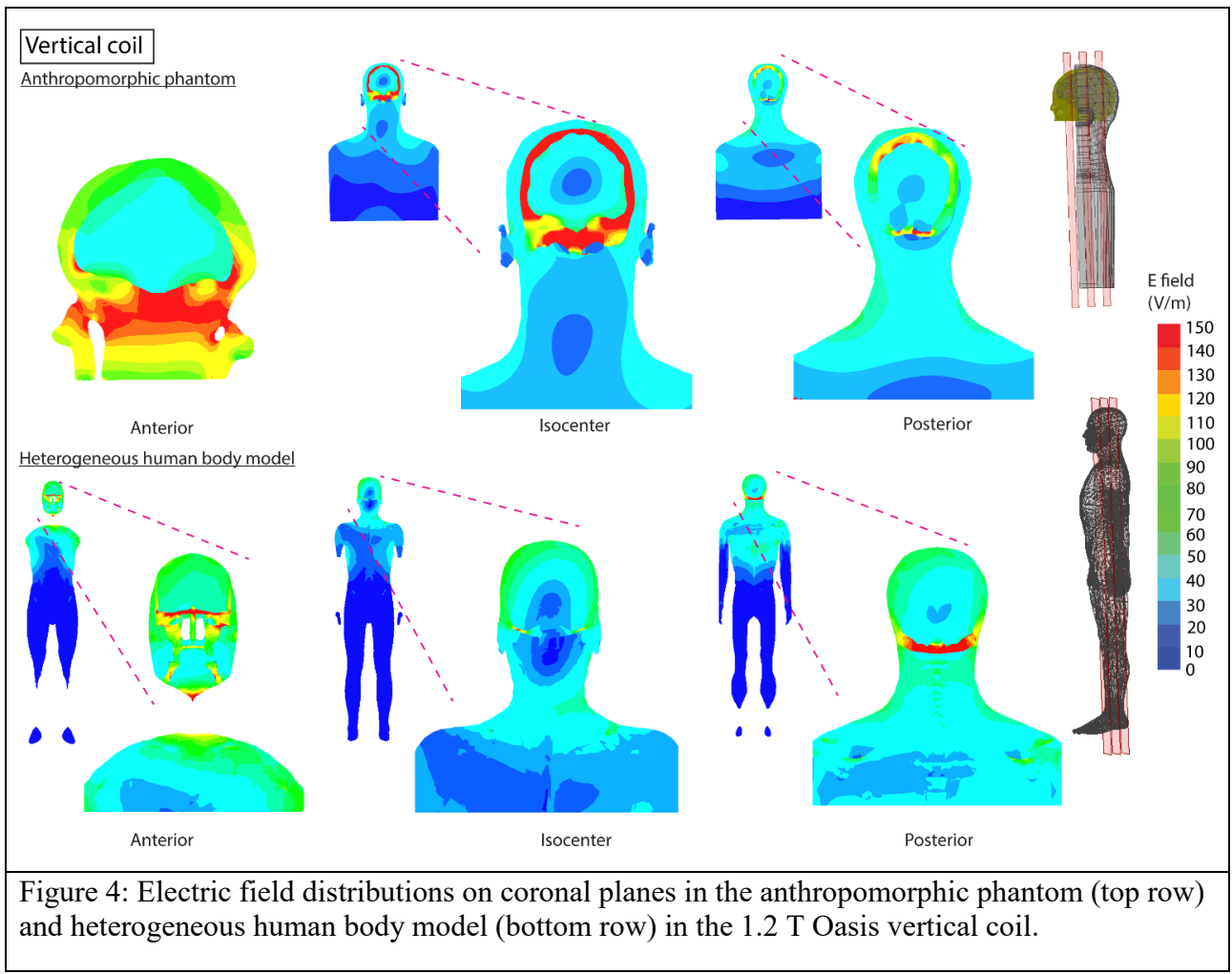

Figure 4: Electric field distributions on coronal planes in the anthropomorphic phantom (top row) and heterogeneous human body model (bottom row) in the 1.2 T Oasis vertical coil.